# Optical and electrical properties of $Nd^{3+}$ doped $Na_2O$-$ZnO$-$TeO_2$ Material

J. N. Mirdda[1], S. Mukhopadhyay[1], K. R. Sahu[2], M. N. Goswami[3*]

[1]J. N. Mirdda: Dept. of Physics, Jadavpur University, Kolkata, 700032, India, email: jyotinessbcollege@gmail.com

[1]S. Mukhopadhyay: Dept. of Physics, Jadavpur University, Kolkata, 700032, India, email: phy.smukherjee@gmail.com

[2]K. R. Sahu: Dept. of Physics, Bhatter College, Paschim Midnapore, 721426, India, email: kriti.basis2020@gmail.com

[3]M. N. Goswami: Dept. of Physics, Midnapore College, Paschim Midnapore, 721101, India, *Email: makhanlal@gmail.com

**Abstract**

Neodymium doped $Na_2O$-$ZnO$-$TeO_2$ (NZT) glasses were prepared by the conventional melt quenching technique. DTA and TG were used to confirmation of glass preparation through the glass transition temperature at 447°C for the glass system. The analysis of FTIR spectra and X-ray diffraction described the nature of the samples were ionic and amorphous respectively. The optical bandgap energy was estimated using absorption spectra and found to be decreased from 2.63eV to 1.32 eV due to the increase of doping concentration. The intensity of the emission spectra was enhanced for the higher concentration of $Nd^{3+}$ ions. The dielectric constant of the glass samples was found to be constant for the large range of frequency (3 kHz to 1 MHz). The variation of conductivity with the temperature of the samples had shown the Arrhenius mechanism of conduction.

Corresponding author: Makhanlal Nanda Goswami, Dept. of Physics, Midnapore College, Paschim Midnapore, West Bengal, India (Email: makhanlal@gmail.com, Mob: 9732730573)

# 1. Introduction

Tellurite glasses are extremely attractive materials for linear and non-linear application in optics, due to their important aspects such as their low melting temperature, low phonon energy, and high refractive index, high dielectric constant, good chemical durability, high thermal stability, non-hygroscopic, with a large transmission window and the possibility to integrate a large amount of rare-earth ions [1-5]. It can be used as micro-lenses, IC photo masking glass, hard disks, press modeling of spherical lenses, glass substrates for solar cells, artificial bones, dental implants, and crown. The optical property of rare-earth ions in tellurite glasses depends on the chemical composition, which determines the structure and the nature of the bonds of the glass matrix. Besides, the understanding of their microscopic mechanism of structural and optical behavior gave much thrust and basic interests for both academia and the industries. Tellurite glasses doped with rare-earth ions have attracted researchers for their broad spectrum of applications in optoelectronic and photo-electronic devices viz solid-state lasers optical switches, broad-band amplifications, nonlinear optical devices, infra-red (NIR) laser windows, optical fibers. The doping of rare-earth ions in tellurite glasses have shown interesting properties like amplification of optical signal in the visible and NIR region, optical data storage, white light emission and planner waveguides which are applicable to micro-chip lasers, biomedical diagnostics, light-emitting diodes and high-density optical data reading [6-8]. Recently, the precise properties of tellurite glass have been investigated for the demonstration of various spectroscopic and nonlinear optical device applications as broadband light emission and optical communications network.

The applications of neodymium-doped glass materials are most commonly useful than the different types of rare-earth ions doped glasses [9]. In 1961, Snitzer initiated the application of glass material as a medium for containing neodymium ions ($Nd^{3+}$) [10]. Neodymium doped lasers have been used in various applications due to perform within high-efficiency range at room temperature. The phonon energy is decreased due to the amalgamation of heavy metal oxides into the tellurite glass system. For this result, thermal stability of glasses is increased [11]. The fundamental structural units of tellurite glass are adapted to the rare-earth oxides like $Nd_2O_3$ from 4-coordination to 3-coordination by exchanging $TeO_4$ trigonal-bipyramid units to the $TeO_3$ trigonal pyramid units [12].

In this present work, the focus has been initiated to study systematically the glass formation, structural, optical and electrical properties constant at room temperature with frequency variation and thermal conduction mechanism of the $Nd^{3+}$ doped $Na_2O$-$ZnO$-$TeO_2$ (NZT) glass materials with different doping concentrations.

## 2. Experimental

The conventional melt quenching technique has been used to prepare $Na_2O$-$ZnO$-$TeO_2$ (NZT) glasses using research great initial ingredients Zinc Oxide (ZnO), Tellurium di-Oxide ($TeO_2$), Sodium Carbonate ($Na_2CO_3$) manufactured by Merck and Neodymium Oxide ($Nd_2O_3$) made by Loba Chemie. The mixing ratio of $Na_2O$, ZnO, $TeO_2$ is maintained as 1:2:7 to prepare the host glass. Neodymium Oxide ($Nd_2O_3$) was added to the host glass as a dopant for (0-2) wt%. The homogeneous mixture of these compounds was obtained by grinding the ingredient powders in an agate mortar. The mixture was kept in an alumina crucible and the crucible was placed in an electrical box furnace. The melt quenching process was obtained by two stages of heating with the temperature at 400 °C for 1 hour and temperature at 475 °C for the next half an hour to produce the quality telluride glass. The cylindrical stainless steel plate was used to hold the melted sample for quenching and the prepared glass was placed again in the furnace at 400°C for 1hour to anneal the sample. The annealed glass was allowed to reach room temperature gradually through the slow cooling process for avoiding thermal stress.

The thermal properties, differential thermal analysis and thermo-gravimetry of the initial mixtures (raw materials in the powder form)were analyzed in the argon environment by using PerkinElmer Instrument (Pyris Diamond TG/DTA, thermo-gravimetric/differential thermal analyzer).These were studied properties for the temperature range of 30 °C to 650 °C with a scanning rate of 10°C/min. X-ray diffraction patterns were obtained using an X-ray diffractometer (RIGAKU model: Japan, XRD 6000, $\lambda$ = 1.5418 Å) with a slow scanning rate 3°/min between the angle 10º and 70ºfor all the samples. FTIR spectrometer (HITACHI Model F-700) was used to identify the nature of pure and doped glasses in the wave number range 400-3000 $cm^{-1}$. Optical absorption and emission spectra of all the glasses were obtained using UV/VIS/NIR spectrophotometer (PerkinElmer Lambda-35) for the wavelength range 400-800 nm and fluorescence spectrophotometer (HITACHI Model F-7000) for 300-600 nm at room temperature. The dielectric constants ($\varepsilon$) of prepared glass samples were measured using LCR-HiTESTER (HIOKI, Japan) for the large frequency range of 200 Hz-4MHz at room temperature. The temperature-dependent DC conductivity of these glasses was studied for the temperature range 36 °C-227 °C using constant voltage supply and current meter.

## 3. Results and Discussion:

The color of the pure NZT glass is white, while $Nd^{3+}$ doped NZT glasses are revolved into purple color due to the doping of $Nd^{3+}$ ions in the host glass NZT. The prepared transparent NZT glasses were shown in Fig. 1. The color of the glass samples varies from bluish to violet with the increase of doping concentration. It has been found that there are no visible crystallites present in these transparent samples [13].

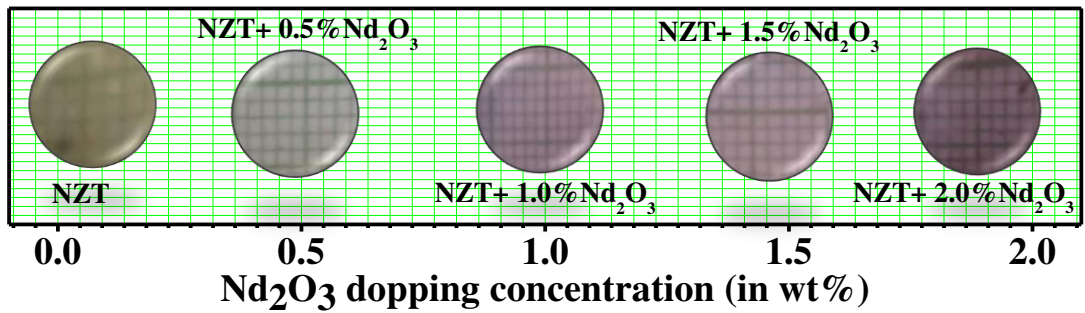

Fig. 1. Pictorial view of $Nd_2O_3$ dipping NZT glasses different samples.

### 3.1. Thermal analysis

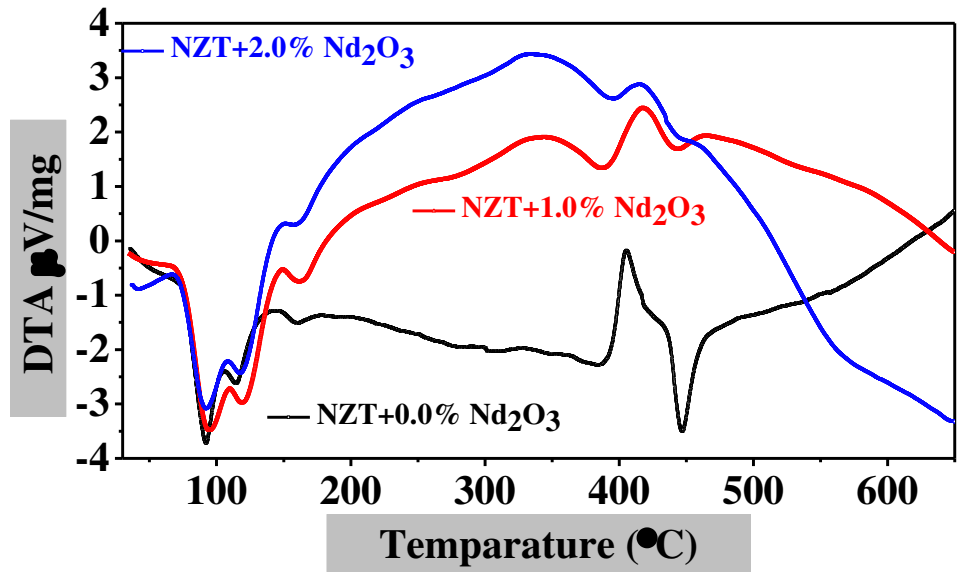

Fig. 2. DTA graphs of pure, 1.0 wt% and 2.0 wt% of $Nd_2O_3$ doped with NZT glasses.

Fig. 2 shows the DTA curve of the mixed precursor of pure and $Nd^{3+}$ ions doped NZT glasses for the temperature range 35 °C to 650 °C. There are three successive endothermic peaks evolved in the temperature range of 72 °C to 160 °C for all the samples. These endothermic peaks in this specified temperature range reveal the desorption or evaporation of the moisture from the precursor ingredients [14]. It has been also observed that there are exothermic peaks present in the DTA curve within the temperature range 330 - 463 °C. These peaks demonstrate the removal of $CO_2$ through the decomposition of $Na_2CO_3$ present in the initial ingredients and the phase transition from solid powders to liquid form through melting. The phase transition has come about through the melting at the temperature of 447 °C. The melting point of $TeO_2$ (730°C) is reduced to 447 °C due to the existence of $Na_2CO_3$ and ZnO in the initial mixture [15, 16]. It is also observed that the melting point of the mixture slightly decreases with increasing the concentration of $Nd_2O_3$.

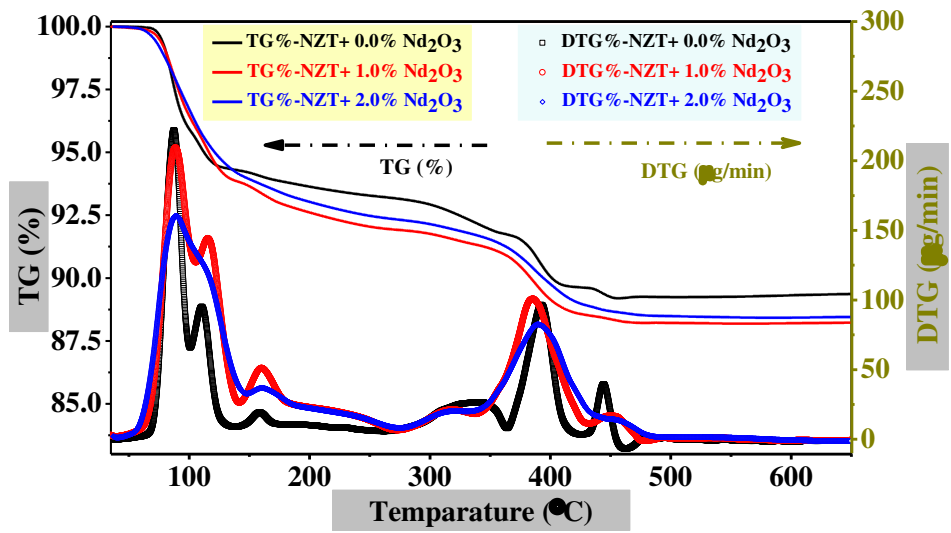

**Fig. 3.** TGA and DTG vs temperature graph of pure, 1.0wt% and 2wt% $Nd_2O_3$ doped with NZT glass samples.

TG and its differentiation curves of the precursor powders of the samples are shown in Fig. 3. These curves for all the samples depict the weight loss during heating at different temperatures. The total weight loss in the measured temperature range 35 °C to 650 °C is 10.71% and this amount of losses occurred from the removal of moisture, evaporation of the volatile substances, and decomposition of $Na_2CO_3$ by releasing $CO_2$, etc. The major mass loss in the temperature range 64-125 °C is taken place due to the desorption of water, removal of moisture and other volatile materials which is also confirmed from DTA curves [17]. The emission of $CO_2$ is observed at 385 °C to 395 °C which is displayed in DTG curves for pure and doped samples. The little amount of mass is reduced within the temperature range of 442 °C to 460°C during the melting of solid powders. No weight loss is observed beyond the temperature of 460 °C.

### 3.2. X-ray Diffraction

Fig. 4 shows the XRD pattern of the glass samples. The XRD peaks of pure NZT glass are observed at 10.86º, 21.92º, 23.26º, 29.03°, 29.86° and 31.39° for the corresponding planes (010), (110), (011), (021), (111) and (030) respectively. It has been found that the sharp peak at 13.24° signifies new hybrid compounds and it may develop due to the presence of $Na_2Te_2O_5, 2H_2O$ [19]. This new hybrid compound has not been formed in case of the addition of doping material $Nd_2O_3$, as the peak position at 13.24° is absent in the doped NZT glass samples. The broadening of XRD peaks in higher doping concentrations reveals the amorphous nature of the doped samples. The shifting of the peak positions is observed for the higher concentration of $Nd_2O_3$ doped NZT glass samples and this shifting has ascribed to the modification of crystalline pattern embedded in the glass samples due to the incorporation of $Nd^{3+}$ ions.

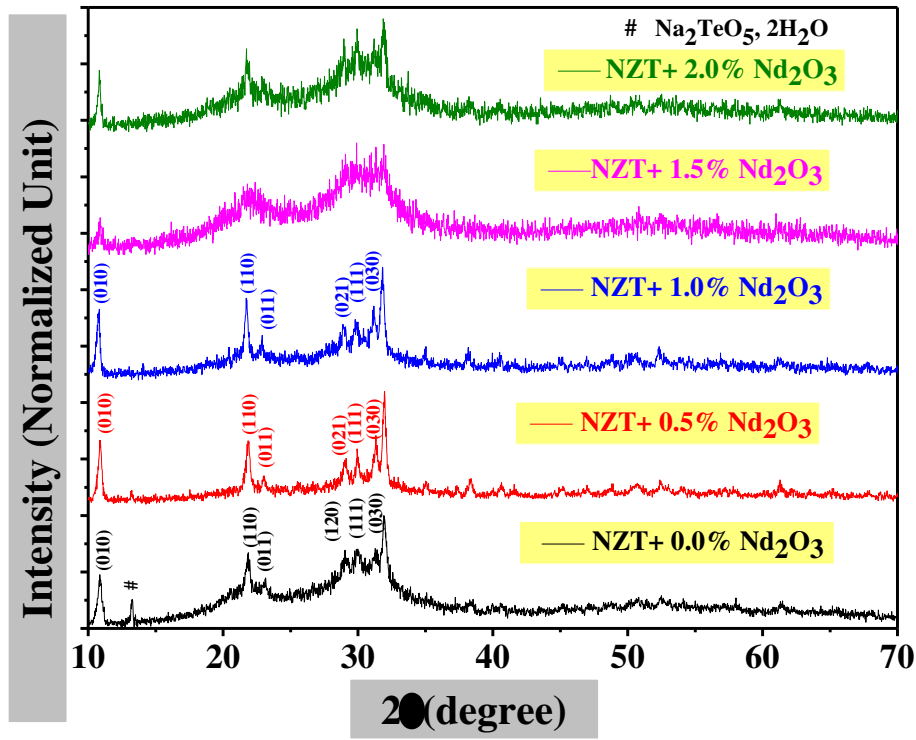

**Fig. 4.** XRD pattern of pure and $Nd^{3+}$ doped with NZT glass samples.

### 3.3. FTIR

The FTIR spectra for pure and doped NZT glasses are shown in Fig. 5. The analysis of these spectra is a useful process to investigate structural studies with functional groups and bonding information in the crystalline and non-crystalline systems [19]. The transmission spectra of the various concentration of neodymium oxide doped glass samples are recorded for the region 400-3000 cm$^{-1}$. The position of the structural unit of ZnO is observed in the band range 424–440 cm$^{-1}$. The absorption band at 426 cm$^{-1}$ is appeared due to the symmetric stretching vibration of the Zn-O bond [20]. The characteristics of tellurite oxide found the structural unit in the range 600-800 cm$^{-1}$. In this broad range, the pure $TeO_2$ is characterized by an infrared absorption at around 644 cm$^{-1}$. The formation of tellurite glass contains two types of the fundamental structural units as $TeO_4$ trigonalbipyramidal and trigonal pyramidal. Symmetrical stretching vibration of Te–O bond in trigonalbipyramids ($TeO_4$) and Te–O bending vibrations in trigonal pyramids ($TeO_3$) in the tellurium network were observed around 644 cm$^{-1}$ [21, 22]. The broad peaks can be attributed to the mixing of two groups and the absorption peaks broaden with the addition of $Nd_2O_3$. This broadenings of peaks and increase of intensity confirm $Nd^{3+}$ ions in the host glass matrices. The band at 771 cm$^{-1}$ is evolved for pure and doped glasses due to trigonal pyramidal structural units. This band is attributed to the stretching vibration within the tellurium and the non-bridging oxygen of trigonal pyramidal structure [23].

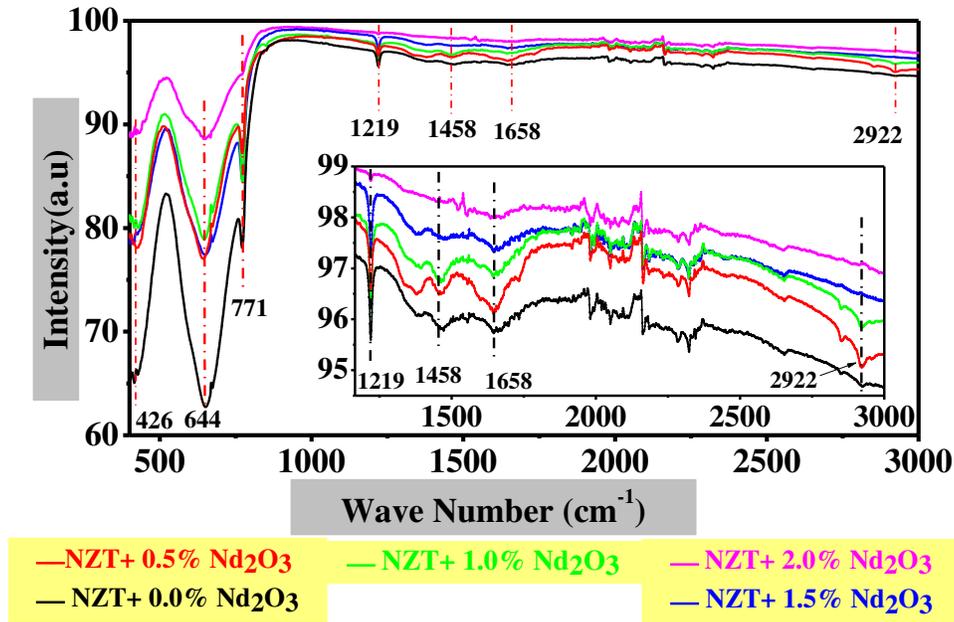

**Fig. 5.** FTIR spectroscopy of pure and different concentrations of $Nd_2O_3$ doped with NZT glass materials.

The absorption peaks around 1658 cm$^{-1}$ and 2922 cm$^{-1}$ are ascribed a stretching vibration of hydrogen bonding and mental bonding with hydroxyl group as the hydroxyl group is coupled with $Te^{4+}$ glass network [24].

**Optical Properties**

**3.3.1. Optical absorption**

The absorption spectra of pure and $Nd_2O_3$-doped NZT glasses are shown in Fig. 6 for the visible region (400-800 nm) at room temperature. It has been found that the absorption transition occurs only for $Nd^{3+}$ ions doped glasses and the intensity of absorption spectra increases with the increase of doping concentrations. There is no transition of the pure glass sample. The peak positions of absorption spectra for $Nd^{3+}$ ions doped glasses are designated as similar to $^4I_{9/2}$–$^4F_J$(J= 9/2,7/2) and $^4I_{9/2}$–$^4G_J$(J=11/2,9/2,7/2,5/2) transitions corresponding to the six absorption band at 430 nm ($^4I_{9/2}$–$^4G_{11/2}$), at 512 nm ($^4I_{9/2}$–$^4G_{9/2}$), at 525 nm ($^4I_{9/2}$–$^4G_{7/2}$), at 583 nm ($^4I_{9/2}$–$^4G_{5/2}$), at 683 nm ($^4I_{9/2}$–$^4F_{9/2}$) and 746 nm ($^4I_{9/2}$–$^4F_{7/2}$). Though the absorption band of rare-earth doped tellurite glasses arises in the ultraviolet region in general, the current glass network displays the absorption in the visible region with large intensity due to the presence of $Nd^{3+}$ ions as a dopant in the glass samples. The introduction of $Nd^{3+}$ ions behaves as crystalline material confirming to the sharp absorption band which is also confirmed from XRD. The different forbidden transitions concerning 4f levels are also involved for the exhibition of absorption bands in the visible region [11]. It has been also found from Fig. 6 that the peak positions of absorption bands for different dopant concentration of $Nd_2O_3$ is slightly shifted towards lower wavelength. This displacement of the absorption band may be occurred due to the change of structural arrangement and various fundamental units of the present glass materials. The peak position of the absorption band may also be shifted due to the change of strength in the oxygen bond in the glass materials. Similar absorption spectra are observed for the addition of rare-earth oxide of the same type in the other tellurite glasses [25-27].

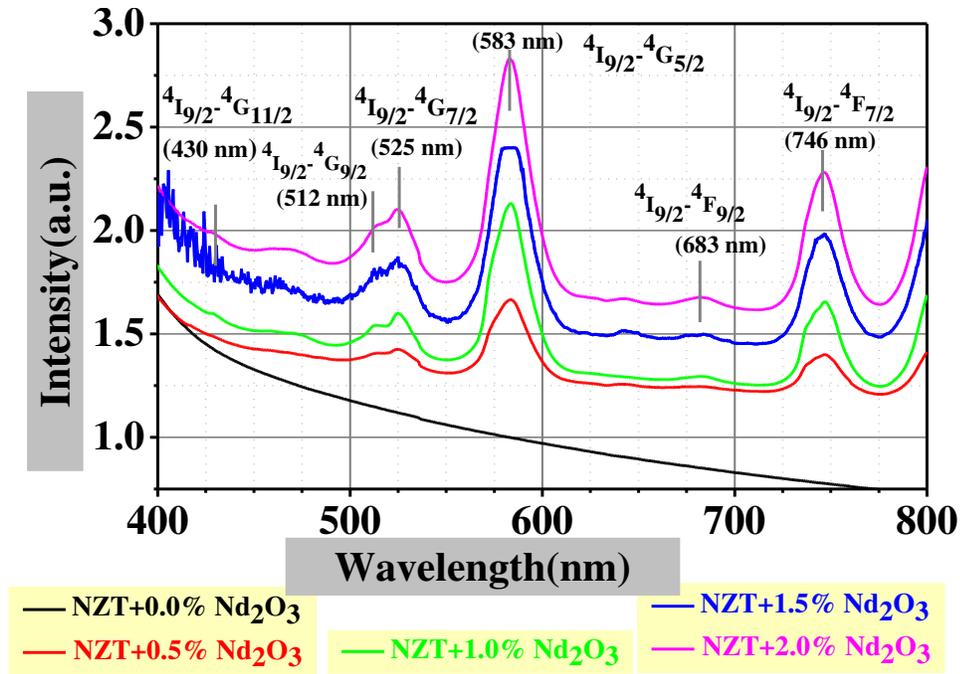

**Fig. 6.** Absorption spectroscopy of pure and $Nd^{3+}$ ions doped with NZT glass materials.

Fig. 7 shows the variation of $(\alpha h\nu)^{1/2}$ with $h\nu$ (Tauc plot) for all of the glass ceramics at room temperature. Here, the coefficient of absorption $\alpha(\upsilon)$ of optical absorption spectra is obtained according to the formula of Davis and Mott

$$\alpha(\nu) = \left(\frac{A}{d}\right) \ldots \ldots \ldots \ldots (1)$$

Where A is the absorbance and d is the thickness of the sample [28, 29].

The indirect bandgap energy is decreased from 2.63 eV and 1.32 eV due to the higher concentration of the $Nd^{3+}$ ions. The bandgap energy and other physical properties of pure and doped glass samples are tabulated in Table 1. The physical properties are performing an important part to estimate the optical property for rare-earth doped glasses. The structural efficiency of the glass samples is developed in the glass network. The physical parameters of prepared glasses like concentration ions, inter-ionic distance ($r_i$), polaron radius ($r_p$) and field strength (F) are calculated and displayed in Table 1. The variation of inter-ionic distance ($r_i$) and field strength (F) with different ion concentrations of rare-earth ion ($Nd_2O_3$) is shown in Fig. 8. It has been found that the Inter-ionic distance ($r_i$) decreases and field strength (F) increases with the increase of doping concentration of $Nd_2O_3$. The Stronger field strength is evolved around $Nd^{3+}$ ions with the increase of Nd-O bond strength. The compactness of the glass structure is confirmed from the result of density and increasing concentration of $Nd^{3+}$ ions in the doped glasses. The electron localization may be formed due to the decrease of $r_i$ and $r_p$. This formation of electron localization can affect directly band gap energy according to the structural change of the glass system. The donor centers in the glass matrix increase with the increase of electron localization and the outcome is the decrease of optical band gap energy ($E_g$) which is also confirmed from the Tauc's plot for pure and $Nd^{3+}$ doped glasses [30-33].

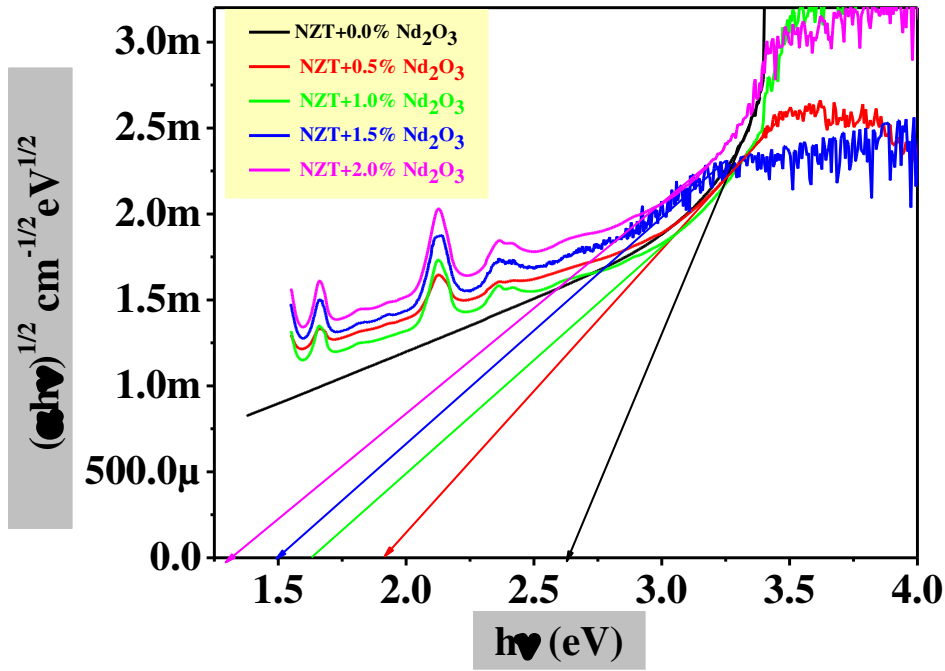

**Fig. 7.** Bandgap energy of $Nd^{3+}$ ions doped with NZT glasses.

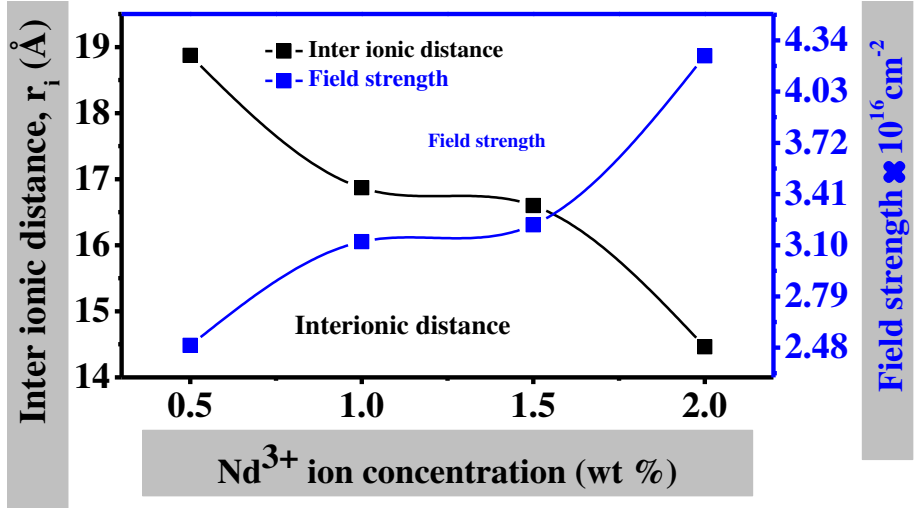

**Fig. 8.** $Nd^{3+}$ ion dependence inter ionic distance ($r_i$) and Field strength (F) for NZT glass compound.

**Table 1** Band gap energy and physical properties of pure and $Nd^{3+}$-doped NZT glass materials.

| Physical properties | $Nd_2O_3$ doped different samples | | | | |
|---|---|---|---|---|---|
| Doping concentration of $Nd_2O_3$ in NZT (wt%) | 0.0 | 0.5 | 1.0 | 1.5 | 2.0 |
| Concentration of $Nd^{3+}$ ions N($10^{20}$ ions/cm$^3$) | - | 1.49 | 2.08 | 2.19 | 3.31 |
| Inter ionic distance between the $Nd^{3+}$ ions ($r_i$ in Å) | - | 18.87 | 16.87 | 16.60 | 14.46 |
| Polaron radius ($r_p$ in Å) | -- | 7.61 | 6.80 | 6.69 | 5.83 |
| Electric Field strength F (in $10^{16}$ cm$^{-2}$) | -- | 2.49 | 3.12 | 3.22 | 4.25 |
| Bandgap Energy $E_g$ indirect (n = 2) eV | 2.63 | 1.90 | 1.63 | 1.50 | 1.32 |

### 3.3.2. Fluorescence Spectra

The fluorescence spectra of pure and $Nd_2O_3$ doped with NZT glasses have shown in Fig. 9. The spectra have been obtained at room temperature for the wavelength range 355nm to 600 nm with an excitation wavelength of 325 nm. No transition takes place for a pure NZT glass sample. The intensity of the emission spectra for all the doped glasses increases with increasing the doping concentration of $Nd^{3+}$ ions. It has been found that a broad spectrum of wavelength 370-556 nm is emitted from the doped samples. This spectrum contains several emission lines which is evaluated from the deconvolution of each spectrum. We have observed five emission bands emitted a broad spectrum $^2P_{3/2} \rightarrow {}^4I_{9/2}$ at 371nm, $^2D_{5/2} \rightarrow {}^4I_{9/2}$ at 421nm, $^2P_{1/2} \rightarrow {}^4I_{9/2}$ at 431 nm, $^4G_{11/2} \rightarrow {}^4I_{9/2}$ at 467 nm, and $^4G_{7/2} \rightarrow {}^4I_{9/2}$ at 546 with the excitation wavelength $\lambda_{exci}$ = 325 nm. The emission corresponding to the $^2P_{1/2} \rightarrow {}^4I_{9/2}$ transition is stronger than the other emissions bands [6, 34]. As the $Nd^{3+}$ ions are populated in the $^2D_{5/2}$ level some of them relax radiatively to this level by emitting the fluorescence. The green emission of $Nd^{3+}$ ion in doped NZT glass is observed due to the decaying transition from $^4G_{7/2} \rightarrow{}^4I_{9/2}$ at 546nm. The peak position of the transitions shifted towards a higher wavelength for 2% doping of concentration of $Nd_2O_3$. This shifting may be arises due to the presence of excited $Nd^{3+}$ ions in $^2D_{5/2}$ and $^4G_{11/2}$ state for the respective transitions with blue emissions [6].

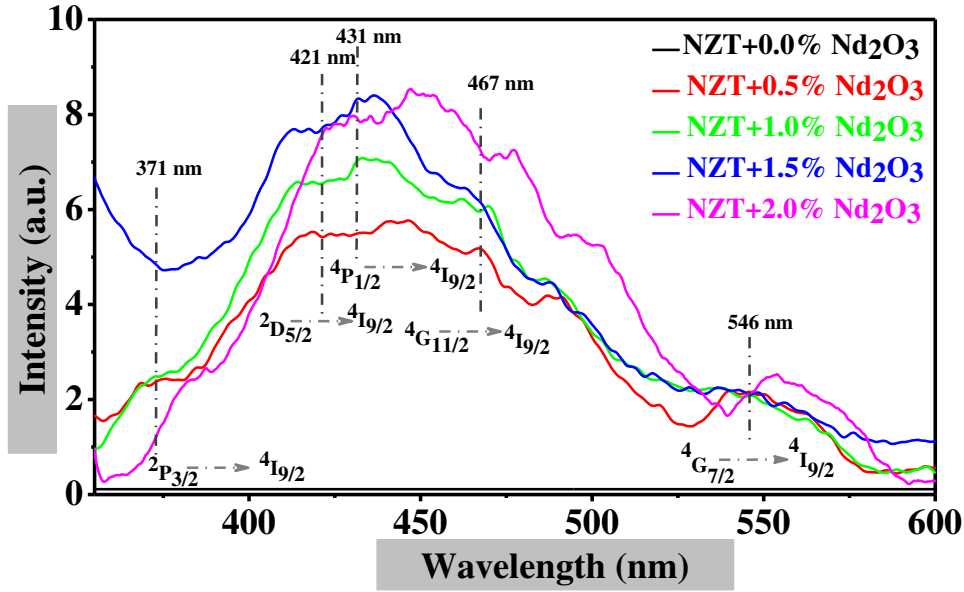

**Fig. 9.** Fluorescence spectra of various concentrations of $Nd^{3+}$ions doped with NZT glasses, excited at 325nm.

### 3.3.3. Cross Section

The absorption cross-section of the glass samples for the $^4G_{5/2}-{}^4I_{9/2}$ transition have been calculated using Lambert-Beer formula

$$\sigma_{ab}(\lambda) = 2.303\ A/(Nd) \quad \ldots \quad \ldots \quad \ldots \quad (2)$$

Where A is absorbance, d is the thickness of the sample and $N$ is the density (ions/cm$^3$) of $Nd^{3+}$ ion of the telluride glass. Fig. 10 shows the absorption cross-section is decreasing for increasing the concentration of $Nd^{3+}$ doped NZT glass samples. The emission cross-section is measured using MaCumber (1964) theory [35].

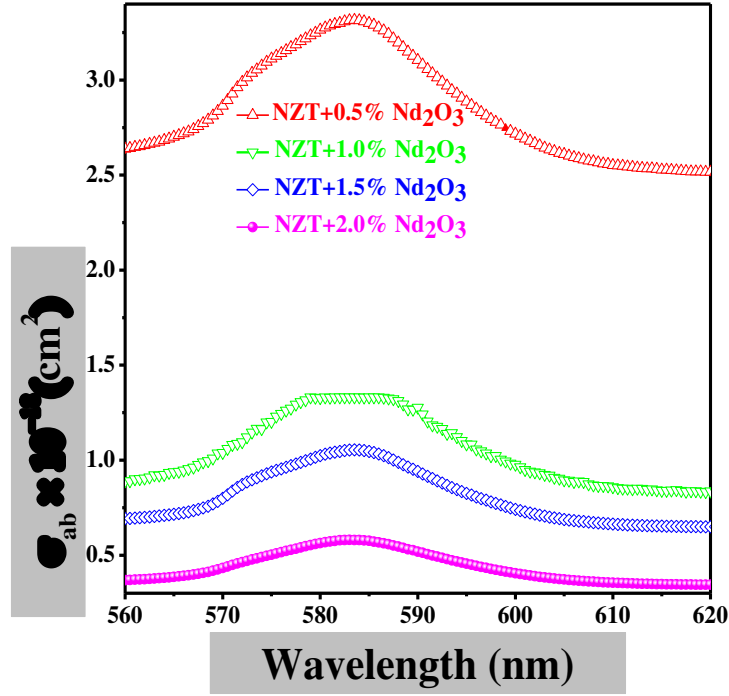

**Fig. 10.** Absorption cross-section vs wavelength (nm) for 0.5 to 2 wt% doped $Nd_2O_3$ with NZT glass materials.

The absorption and emission cross-section are associated with the formula

$$\sigma_{em}(\lambda) = \sigma_{ab}(\lambda)\exp[(E-h\nu)/kT] \quad \ldots \quad \ldots \quad \ldots \quad (3)$$

Where $\nu$ be the phonon frequency, $E$ is the free energy need to excite $Nd^{3+}$ from $^4I_{9/2} \rightarrow {}^4G_{5/2}$ state at temperature $T$, $h$ is the Planck's constant, and $k$ is the Boltzmann constant. The emission cross section for the $^4I_{9/2} \rightarrow {}^4G_{5/2}$ transition of $Nd^{3+}$ doped glasses are shown in Fig.11. The emission cross-section is decreased for a higher concentration of $Nd^{3+}$ ions of NZT glasses.

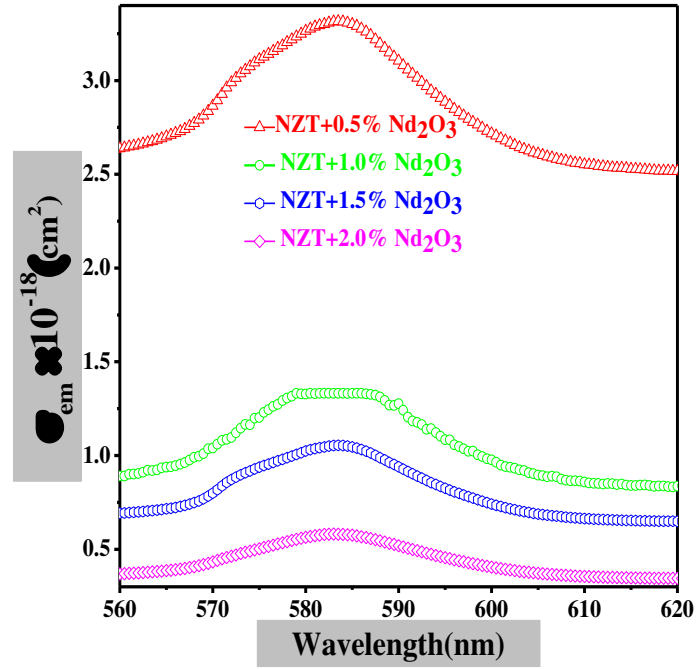

**Fig. 11.** Wavelength (nm) dependence emission cross-section for $Nd_2O_3$ doped with NZT glass materials.

### 3.4. Dielectric Property

### 3.4.1. Dielectric constant

Fig. 12 shows the real part of the dielectric constant (ε′) of $Nd^{3+}$ doped NZT at room temperature in the frequency range 200 Hz to 4MHz. The value of ε′ of $Nd^{3+}$ doped NZT glass samples is decreased with increasing frequency from 200 Hz to ~ 2 kHz. At lower frequencies, the value of ε′ is high due to the absence of spontaneous polarization in oxide glass materials [36]. The values of ε′ of our samples are almost frequency independent in the frequency range ~ 3kHz to ~ 1MHz due to decrease of ionic, space charge, and orientation polarization. The dielectric constant is increased in the frequency range ~ 1 MHz to 4 MHz with higher concentration of $Nd^{3+}$ ions. A similar result has been observed in $Eu^{3+}$ ions doped NZT glass samples [37].

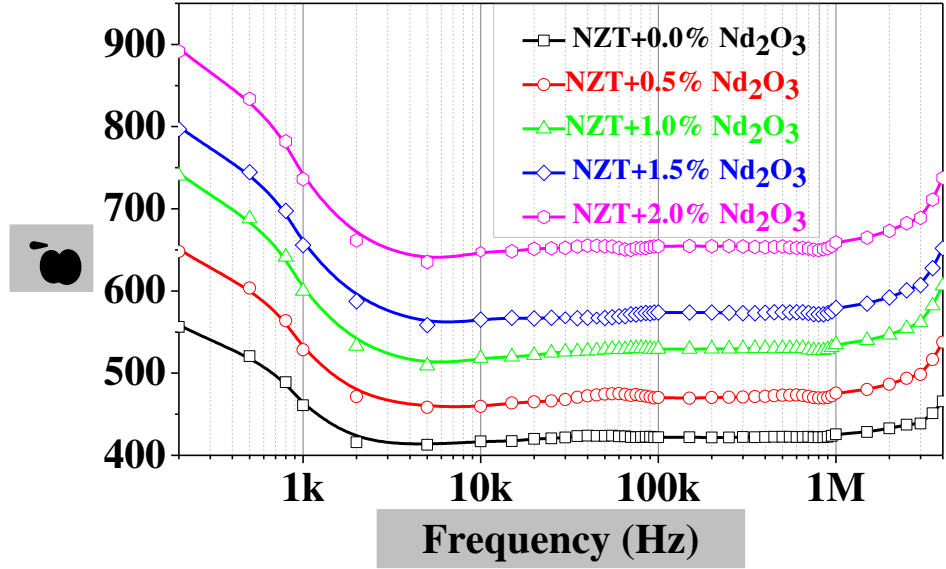

**Fig. 12.** The variation of dielectric constant (ε′) as a function of frequency for $Nd^{3+}$ doped NZT glasses.

### 3.4.2. DC Conductivity

Fig. 13 shows the variation of electrical conductivity with temperature for pure and $Nd^{3+}$ doped NZT glass materials. It has been found that the conductivity of all the glass samples is increased with increasing temperature and also with the increase of doping concentration in the host glass. This increment of conductivity with temperature reveals that the electrical conduction mechanism is Arrhenius type. This Arrhenius mechanism of electrical conduction can be used and estimated activation energy of the samples using the relation

$$\sigma_{dc} = \sigma_0 \exp\left(-\frac{E_a}{kT}\right) \quad \ldots\ldots (2)$$

Where $E_a$ is the activation energy, $\sigma_0$ is the pre-exponential factor, $T$ is the absolute temperature and $k$ is the Boltzmann constant. The estimated activation energies for all the samples have been calculated from Fig. 13. The activation energy decreases (666 meV to 552 meV) with increasing concentrations of $Nd_2O_3$ doped glass samples [36].

The increment of conductivity due to the increase of doping concentration of rare-earth ions can be explained based on interaction between the rare-earth ions and structural units of host glass. The atomic weight, doping concentration and location of $Nd^{3+}$ ions describe the mechanism of conductivity in the glass structure. The addition of rare-earth ions in the host glass creates a large number of non-bridging oxygen in the glass materials. This is also confirmed from the analysis of obtained FTIR spectra for all the samples. So, the increases of $Nd^{3+}$ ions in the NZT glasses enhance the creation of non-bridging oxygen atoms, and hence the conductivity is increased for all the doped glass samples [38].

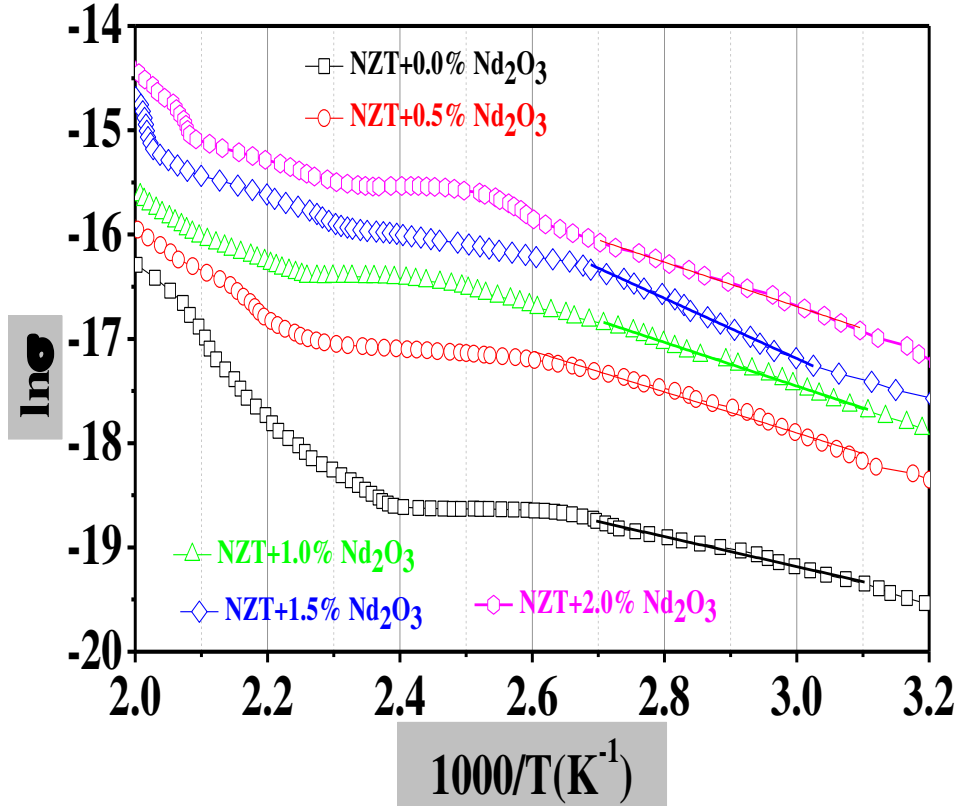

**Fig. 13.** Arrhenius plot of $\ln\sigma$ vs 1000/T for $Nd^{3+}$ ions doped NZT glasses.

**Conclusions:**

Neodymium (III) doped NZT glasses were effectively manufactured by the melt quenching method. The X-ray diffraction characterization sturdily signified the amorphous nature of the prepared glasses for the higher concentration of $Nd^{3+}$ ions. The results of the preparation and characterization of the luminescent system are obtained for the addition of various concentrations (0.5 wt%, 1.0 wt%, 1.5 wt%, 2.0 wt%) of NZT glass samples. Six absorption bands are displayed corresponding to 430 nm ($^4I_{9/2}$–$^4G_{11/2}$), at 512 nm ($^4I_{9/2}$–$^4G_{9/2}$), at 525 nm ($^4I_{9/2}$–$^4G_{7/2}$), at 583 nm ($^4I_{9/2}$–$^4G_{5/2}$), at 683 nm ($^4I_{9/2}$–$^4F_{9/2}$), and 746 nm ($^4I_{9/2}$–$^4F_{7/2}$). The peak value of photoluminescence spectra is slightly changed due to the addition of the higher concentration of rare-earth ion of NZT glass sample. It specifies that the excitation wavelength simply depends upon the glass composition. Field strength (F) was observed to increase with increasing the $Nd^{3+}$ ions concentration. Inter ionic distance ($r_i$) and polaron radius ($r_p$) were found to decrease with increasing the $Nd^{3+}$ ions. The role of $Nd^{3+}$ ions in modifying the structural and optical properties has been understood. The decrease in the value of direct optical bandgap energy was ascribed to the change in the structure of such glasses. The dielectric constant of $Nd_2O_3$ doped NZT glasses was measured for various frequencies and seen as a stable substance within the frequency of 3 kHz to 1 MHz. So $Nd_2O_3$ doped NZT glass samples have a large prospect for the application in optoelectronic devices.

**Acknowledgement:**

The authors are grateful to the CRF, IIT Kharagpur for providing facilities to study DTA and TGA. The authors desire to express thanks to Jadavpur University for providing facilities to study FTIR. The work is partly supported by DST Govt. of West Bengal research project (Memo No.: 296 (Sanc)/ST/P/S&T/16G- 17/2017) of India and also thanks to all authors of the useful references.